\documentstyle[aps,twocolumn,pra,psfig]{revtex}

\begin{document}
\title{Casimir force between metallic mirrors}
\author{Astrid Lambrecht and Serge Reynaud}
\address{Laboratoire Kastler Brossel \thanks{%
Unit\'{e} de l'Ecole Normale Sup\'{e}rieure, de l'Universit\'{e} Pierre
et Marie Curie, et du Centre National de la Recherche Scientifique.}, \\
Campus Jussieu, case 74, \\
75252 Paris Cedex 05, France}
\date{September 1999}
\maketitle

\begin{abstract}
We study the influence of finite conductivity of metals on the Casimir effect. 
We put the emphasis on explicit theoretical evaluations
which can help comparing experimental results with theory. The reduction of
the Casimir force is evaluated for plane metallic plates. The reduction of
the Casimir energy in the same configuration is also calculated. It can be
used to infer the reduction of the force in the plane-sphere geometry
through the `proximity theorem'. Frequency dependent dielectric response
functions of the metals are represented either by the simple plasma model
or, more accurately, by using the optical data known for the metals used in
recent experiments, that is Al, Au and Cu. In the two latter cases, the
results obtained here differ significantly from those published recently.

\end{abstract}

\section{Introduction}

The Casimir force experienced by reflectors placed in vacuum is a
macroscopic mechanical consequence of quantum fluctuations of
electromagnetic fields \cite{Casimir48}. Despite its relatively small
magnitude, it has been observed in a number of `historic' experiments \cite
{Deriagin57,Sparnaay,Tabor68,Sabisky73}. A much better experimental
precision has been reached in recent experiments \cite
{Lamoreaux97,Mohideen98} which should now allow for an accurate comparison
with theory. Clearly this requires not only a detailed control of the
experiments but also a careful theoretical estimation of the various
corrections corresponding to the differences between real experiments and
the idealized Casimir situation.

The present paper is focussed on the estimation of corrections associated
with the non ideal behavior of metallic reflectors. Additional corrections
due to the effect of non-zero temperature and the geometry of the cavity
have also to be mastered before an agreement of experimental results with
theoretical expectations can be claimed. A general discussion of the
corrections to Casimir formulas is presented in the next section. In
particular, we recall how the Casimir force measured in the plane-sphere
geometry may be inferred from the Casimir energy in the plane-plane geometry
by using the so-called `proximity theorem'.

We then focus attention on our main topic which is the evaluation of the
reduction factor of Casimir force and Casimir energy for plane metallic
plates in the limit of a large surface. We first compute the reduction
factors obtained when describing the dielectric functions with a plasma
model. This computation covers the whole range of distances large or small
with respect to the plasma wavelength. The analytical expression of the
force in the limit of small distances is also derived.

The plasma model is not a good description of the dielectric constant at low
frequencies because it ignores the relaxation of electrons responsible for
optical response of metals. This is why we also investigate the Drude model
which accounts for this relaxation. We finally discuss in detail a more
accurate description of the dielectric constant based on the optical data
known for the metals. We concentrate on the three metals, Aluminium, Gold
and Copper, used in recent experiments and give the reduction factors for
the whole range of experimentally explored distances. For Au and Cu, we
obtain results differing significantly from recently published ones \cite
{Lamoreaux99}.

\section{Corrections to the Casimir formula}

In the original point of view \cite{Casimir48}, the Casimir effect is
derived from the change of the total energy of vacuum due to the presence of
two plane perfect reflectors. In this global approach, the Casimir energy is
the part $E_{C}$ of vacuum energy depending on the plate separation $L$ 
\begin{equation}
E_{C}=A\frac{\hbar c\pi ^{2}}{720L^{3}}  \label{CasimirEnergy}
\end{equation}
This energy is proportional to the surface $A$ of the reflectors in the
limit of a large surface, the Planck constant $\hbar $ and the speed of
light $c$. The Casimir force $F_{C}$ between the two reflectors is then
derived from this position dependent energy 
\begin{equation}
F_{C}=-\frac{{\rm d}E_{C}}{{\rm d}L}=A\frac{\hbar c\pi ^{2}}{240L^{4}}
\label{CasimirForce}
\end{equation}
Being proportional to the surface, it defines a pressure which depends only
on the distance $L$ and the two fundamental constants $\hbar $ and $c$.
Conventions of sign have been chosen so that both numbers (\ref
{CasimirEnergy},\ref{CasimirForce}) are positive with the significance that
the Casimir force is attractive and the Casimir energy is a binding energy.

In contrast with this global point of view, the Casimir force may also be
understood as a local quantity, namely the radiation pressure exerted upon
mirrors by vacuum fluctuations which are modified by the presence of the
reflectors. This local approach makes it much easier to deal with
corrections of Casimir formulas (\ref{CasimirEnergy},\ref{CasimirForce}). In
a remarkable work, Lifshitz gave a general formula for the Casimir force
between two plane plates characterized by their dielectric response
functions \cite{Lifshitz}. In particular his formula accounts for the finite
reflectivity of metallic mirrors and it was used to deduce a first order
correction for the plasma model of metals \cite{Dzyaloshinshii61}. Since the
dielectric constant is large at frequencies smaller than the plasma
frequency $\omega _{P}$, the Casimir formula is recovered at distances
larger than the plasma wavelength 
\begin{equation}
\lambda _{P}=\frac{2\pi c}{\omega _{p}}
\end{equation}
At frequencies larger than $\omega _{P}$ in contrast, the mirror has a poor
reflectivity so that the force is reduced with respect to (\ref{CasimirForce}%
) at distances of the order of the plasma wavelength $\lambda _{P}$ which
lies in the sub-$\mu $m range. The force may thus be written in terms of a
factor $\eta _{F}$ which measures the reduction of the force with respect to
the case of perfect mirrors 
\begin{equation}
F=\eta _{F}F_{C}  \label{defetaF}
\end{equation}
The expression of the reduction factor is read at long distances as \cite
{Hargreaves65,Schwinger78} 
\begin{equation}
\eta _{F}\simeq 1-\frac{8}{3\pi }\frac{\lambda _{P}}{L}\quad \Leftarrow
\quad \frac{\lambda _{P}}{L}\ll 1  \label{etalongF}
\end{equation}
In the following we will also introduce a reduction factor $\eta _{E}$ for
the Casimir energy $E$ 
\begin{equation}
E=\int_{L}^{\infty }F\left( x\right) ~{\rm d}x=\eta _{E}E_{C}
\label{defetaE}
\end{equation}

The Lifshitz formula also contained thermal corrections to the Casimir
effect usually studied at zero temperature. These corrections are
significant at distances larger than or of the order of a typical length 
\cite{Lifshitz} 
\begin{equation}
\lambda _{T}=\frac{\hbar c}{k_{B}T}
\end{equation}
where $k_{B}$ is the Boltzmann constant and $T$ the temperature. At room
temperature this length is of the order of a few $\mu $m. Hence thermal
corrections become significant for distances for which the mirrors can be
considered as nearly perfect reflectors. Well established estimations of
thermal corrections for perfect mirrors may thus be used to evaluate the
effect of temperature \cite{Schwinger78,Mehra67,Brown69}. This effect is
found to be negligible at distances smaller than 1 $\mu $m and to become
dominant at distances larger than 5 $\mu $m \cite{Mohideen98,Klimchitskaya99}.

The Lifshitz formula has been derived for plane plates in the limit of large
transverse surface and large longitudinal optical depth. But recent
experiments have been performed in a plane-sphere geometry which makes
easier the control of geometry and, in particular, the precise control of
the distance between plates \cite{Lamoreaux97,Mohideen98}. Furthermore,
mirrors are often built as multilayered structures rather than as single
plates with a large optical depth. Finally the roughness of the metal/vacuum
interfaces may also play an important role. These features have to be taken
into account in an accurate estimation.

We will not present a detailed analysis of the geometrical effects in this
paper. It is however worth recalling that the Casimir force in the
plane-sphere geometry is usually estimated from the so-called `proximity
theorem'. Basically this theorem amounts to evaluating the force by adding
the contributions of various distances as if they were independent. In the
plane-sphere geometry the force $F_{pt}$ evaluated from the proximity
theorem is thus read as \cite{Deriagin68,Blocki77,Mostepanenko85,Bezerra97} 
\begin{equation}
F_{pt}=2\pi RE=2\pi R\eta _{E}E_{C}
\end{equation}
where $R$ is the radius of the sphere and $E$ the Casimir energy evaluated
in the plane-plane configuration for the same distance $L$, this distance
being defined as the distance of closest approach in the plane-sphere
geometry. Hence, the reduction factor $\eta _{E}$ for the Casimir energy
evaluated in the plane-plane configuration in (\ref{defetaE}) can be used to
infer the reduction factor for the force measured in the plane-sphere
geometry through the proximity theorem.

At this point we have to emphasize that our calculations are intended to
provide a reliable estimation of the Casimir force and energy between two
metal plates in the plane-plane geometry. Clearly, they do not give any
indication of the degree of reliability of the proximity theorem. Since it
is well known that the Casimir force is not an additive quantity one cannot
but question an estimation based on an addition procedure \cite{Hagen99}.
Precisely, one can hardly admit that the proximity theorem provides reliable
estimations at the level of accuracy which is now aimed at, that is the \%
level. As already discussed, these problems are not the main task of this
paper which is focussed on the effect of imperfect reflection of metallic
mirrors. The same remarks apply to another aspect of the geometry, that is
the roughness effect which has also been found to play a significant role 
\cite{Bezerra97,Roy99PRL}.

\section{The description of mirrors}

As predicted by Casimir in his founding article \cite{Casimir48} the
divergences associated with the infiniteness of vacuum energy do not play a
real role in the estimation of Casimir effect thanks to a general physical
reason: real mirrors are certainly transparent at the limit of infinite
frequencies. This idea was implemented in the Lifshitz theory \cite{Lifshitz}
and it has a much broader range of validity. Real mirrors may always be
characterized by frequency dependent reflectivity amplitudes which provide a
finite expression for Casimir energy as soon as general properties of
unitarity, causality and high-frequency transparency are accounted for \cite
{Jaekel91}. Dispersive optical response functions necessarily include
dissipation mechanisms so that incoming electromagnetic fields and
additional Langevin fluctuations coming from matter have to be treated
simultaneously \cite{Kupiszewska92}. The description of mirrors through well
behaved reflectivity amplitudes \cite{Jaekel91} automatically includes a
proper description of these fluctuations \cite{Barnett96}.

The two mirrors form a Fabry-P\'{e}rot cavity which enhances or decreases
field fluctuations depending on whether their frequency is resonant or not
with cavity modes. This modulation of intracavity energy of vacuum
fluctuations, integrated over frequencies and incidence angles corresponding
to the various modes, is responsible for the Casimir force \cite{Jaekel91}.
Using causality properties, the force can be written as an integral over
imaginary frequencies and wavevectors \cite{Lifshitz}. After these
transformations, the Casimir force may be written in terms of a reduction
factor (\ref{defetaF}) which takes the following form adapted from \cite
{Jaekel91} 
\begin{eqnarray}
&&\eta _{F}=\frac{120}{\pi ^{4}}\int_{0}^{\infty }{\rm d}KK^{2}\int_{0}^{K}%
{\rm d}\Omega \sum_{p}\frac{r_{p}^{2}}{e^{2K}-r_{p}^{2}}  \nonumber \\
&&K=\kappa L\qquad \Omega =\omega \frac{L}{c}  \label{etaF}
\end{eqnarray}
$r_{p}$ denotes the reflection amplitude for one of the two mirrors and a
given polarization $p$. This notation is a shorthand for $r_{p}\left(
i\omega ,i\kappa \right) $ where $i\omega $ is the imaginary frequency and $%
i\kappa $ the imaginary wavevector along the longitudinal direction of the
Fabry-P\'{e}rot cavity. $\Omega $ and $K$ stand for the frequency and
wavevector measured in dimensionless units with the help of the cavity
length $L$. The reflection amplitudes are supposed to be identical for the
two mirrors. Otherwise $r_{p}^{2}$ has to be replaced by the product of the
two amplitude reflection coefficients.

In the limit of perfect mirrors $\left( r_{p}^{2}=1\right) $ the Casimir
formula (\ref{CasimirForce}) is recovered $\left( \eta _{F}=1\right) $. In
the general case, the factor $\eta _{F}$ measures the reduction of the force
between real mirrors with respect to the case of perfect mirrors. We may
also write a reduction factor (\ref{defetaE}) for the Casimir energy 
\begin{equation}
\eta _{E}=-\frac{180}{\pi ^{4}}\int_{0}^{\infty }{\rm d}KK\int_{0}^{K}{\rm d}%
\Omega \sum_{p}{\rm Log}\left( 1-r_{p}^{2}e^{-2K}\right)  \label{etaE}
\end{equation}
Let us stress again that the expressions (\ref{etaF},\ref{etaE}) give only
the corrections to Casimir force and energy associated with the finite
conductivity of metallic plates. They correspond to plane reflecting plates
at the limit of a large surface, assume a null temperature and disregard the
problem of roughness. As discussed in the previous section, the factor $\eta
_{E}$ may be used to infer the force in the plane-sphere geometry through
the proximity theorem.

Assuming furthermore that the metal plates have a large optical thickness,
the reflection coefficients $r_{p}$ correspond to the ones of a mere
vacuum-metal interface \cite{LandauECM10} 
\begin{eqnarray}
r_{\bot } &=&-\frac{\sqrt{\omega ^{2}\left( \varepsilon \left( i\omega
\right) -1\right) +c^{2}\kappa ^{2}}-c\kappa }{\sqrt{\omega ^{2}\left(
\varepsilon \left( i\omega \right) -1\right) +c^{2}\kappa ^{2}}+c\kappa } 
\nonumber \\
r_{||} &=&\frac{\sqrt{\omega ^{2}\left( \varepsilon \left( i\omega \right)
-1\right) +c^{2}\kappa ^{2}}-c\kappa \varepsilon \left( i\omega \right) }{%
\sqrt{\omega ^{2}\left( \varepsilon \left( i\omega \right) -1\right)
+c^{2}\kappa ^{2}}+c\kappa \varepsilon \left( i\omega \right) }
\label{rThick}
\end{eqnarray}
$r_{p}$ still stands for $r_{p}\left( i\omega ,i\kappa \right) $ and $%
\varepsilon \left( i\omega \right) $ is the dielectric constant of the metal
evaluated for imaginary frequencies. Taken together, the relations (\ref
{etaF},\ref{rThick}) reproduce the Lifshitz expression for the Casimir force 
\cite{Lifshitz}. We however emphasize that (\ref{etaF}) can be used to go
beyond the Lifshitz expression since it allows one to deal with more general
mirrors than those considered in (\ref{rThick}).

As an example we consider mirrors built as metallic slabs having a finite
thickness. For a given polarization, we denote by $\rho $ the reflection
coefficient (\ref{rThick}) corresponding to a single vacuum/metal interface
and we write the reflection amplitude $r$ for the slab of finite thickness
through a Fabry-P\'{e}rot formula 
\begin{eqnarray}
r &=&\rho \frac{1-e^{-2\delta }}{1-\rho ^{2}e^{-2\delta }}  \nonumber \\
\delta &=&\frac{D}{c}\sqrt{\omega ^{2}\left( \varepsilon \left( i\omega
\right) -1\right) +c^{2}\kappa ^{2}}  \label{rSlab}
\end{eqnarray}
This expression has been written directly for imaginary frequencies. The
parameter $\delta $ represents the optical length in the metallic slab and $%
D $ the physical thickness. The single interface expression (\ref{rThick})
is recovered in the limit of a large optical thickness $\delta \gg 1$. With
the plasma model, this condition just means that the thickness $D$ is larger
than the plasma wavelength $\lambda _{P}$.

In order to discuss recent experiments it may be useful to write the
reflection coefficients for multilayer mirrors. For example one may consider
two-layer mirrors with a layer of thickness $D$ of a metal $A$ deposited on
a large slab of metal $B$ in the limit of large thickness. The reflection
formulas are then obtained as in \cite{Lambrecht97} but accounting for
oblique incidence. The equations (\ref{etaF},\ref{etaE}) may then be
calculated for the two-layer mirrors. This could help to obtain more
accurate estimations for the experiments as soon as physical characteristics
of the two-layer mirrors are precisely known. In the present paper we use
reflection amplitudes (\ref{rThick}) which are well adapted to a general
discussion since they depend on a smaller number of parameters.

\section{Plasma and Drude model}

We will now evaluate the reduction factor for the Casimir force when the
frequency dependent dielectric function may be represented by the plasma or
Drude models.

\begin{figure}[tbp]
\centerline{\psfig{figure=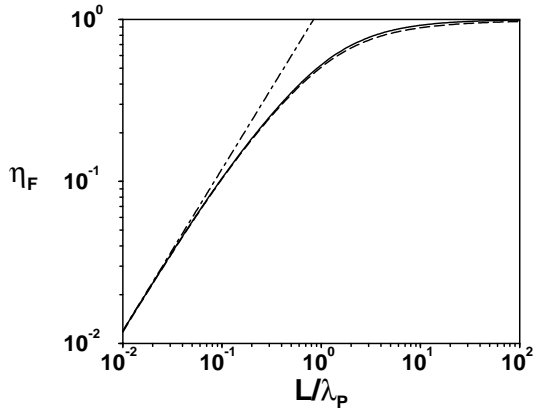,width=7cm}}
\caption{Reduction of the Casimir force compared to the force between
perfect mirrors, when the finite conductivity may be described by a plasma
model (solid line) or a Drude model (dashed line) with a ratio $\frac{\gamma 
}{\omega _{P}}$ equal to $4 \times 10^{-3}$. The difference due to the
relaxation parameter has only a small effect on the calculation of the
Casimir force. The dotted-dashed line corresponds to the short distance
asymptotic behavior.}
\label{fig-plasma}
\end{figure}

We begin with the plasma model where the quantities 
$\varepsilon \left( \omega \right) $ and $\varepsilon \left(
i\omega \right) $ are represented as follows 
\begin{eqnarray}
\varepsilon \left( \omega \right) &=&1-\frac{\omega _{P}^{2}}{\omega ^{2}} 
\nonumber \\
\varepsilon \left( i\omega \right) &=&1+\frac{\omega _{P}^{2}}{\omega ^{2}}
\label{epsPlasma}
\end{eqnarray}
Using expressions (\ref{rThick},\ref{epsPlasma}) it is possible to obtain
the reduction factor (\ref{etaF}) defined for the Casimir force through
numerical integrations. The result is drawn on figure \ref{fig-plasma}, as a
function of the dimensionless parameter $\frac{L}{\lambda _{P}}$, that is
the ratio between the distance $L$ and the plasma wavelength $\lambda _{P}$.
As expected the Casimir formula is reproduced at large distances ($\eta
_{F}\rightarrow 1$ when $\frac{L}{\lambda _{P}}\gg 1$). At distances smaller
than $\lambda _{P}$ in contrast, a significant reduction factor is obtained.
This factor $\eta _{F}$ scales as $\frac{L}{\lambda _{P}}$ at the limit of
small distances. This means that the whole expression (\ref{defetaF}) of the
Casimir force is a power law which undergoes a change of exponent when the
distance $L$ crosses the plasma wavelength $\lambda _{P}$ characterizing the
optical response of metals. This is quite analogous to the crossover
discovered by Casimir and Polder for the variation of Van der Waals force
with respect to the interatomic distance \cite{CasimirPolder}.

An asymptotic law of variation for $\eta _{F}$ varying as $\frac{L}{\lambda
_{P}}$ at small distances has been proposed repeatedly since Lifshitz \cite
{Lifshitz}. We have been able to derive from (\ref{etaF},\ref{rThick},\ref
{epsPlasma}) a precise value for the coefficient appearing in this law 
\begin{eqnarray}
L &\ll &\lambda _{P}\quad \rightarrow \quad \eta _{F}=\alpha \frac{L}{%
\lambda _{P}}  \nonumber \\
\alpha &=&\frac{30}{\pi ^{2}}\int_{0}^{\infty }{\rm d}K~e^{-\frac{3K}{4}%
}\left( \frac{K^{2}}{\sqrt{\sinh \frac{K}{2}}}-\frac{K^{2}}{\sqrt{\cosh 
\frac{K}{2}}}\right)  \nonumber \\
&\simeq &1.193  \label{etashortF}
\end{eqnarray}
This implies a similar behavior for the reduction factor $\eta _{E}$ with a
different proportionality coefficient 
\begin{equation}
L\ll \lambda _{P}\quad \rightarrow \quad \eta _{E}=\frac{3}{2}\alpha \frac{L%
}{\lambda _{P}}  \label{etashortE}
\end{equation}
Hence, $\eta _{E}$ is larger than $\eta _{F}$ at short distances, which
means a less important reduction with respect to the case of perfect
mirrors. The asymptotic law (\ref{etashortF}) valid at short distances,
taken with its equivalent (\ref{etalongF}) at large distances, is an
important feature of the variation of $\eta _{F}$ with $L$ which is not
obeyed by the approximants which have been used to discuss recent
experimental results \cite{Lamoreaux99,Klimchitskaya99}. These approximants
are compared with the exact expression of the reduction factor for the
plasma model in Appendix \ref{app-Plasma}.

As already discussed, the plasma model does not provide a good description
of the dielectric response of metals. The main reason is that the dielectric
function $\varepsilon \left( \omega \right)$ is real in (\ref{epsPlasma})
and, therefore, does not account for any dissipative mechanism. A much
better representation of the dielectric function corresponding to the
optical response of conduction electrons is the Drude model \cite
{AshcroftMermin} 
\begin{eqnarray}
\varepsilon \left( \omega \right) &=&1-\frac{\omega _{P}^{2}}{\omega \left(
\omega +i\gamma \right) }  \nonumber \\
\varepsilon \left( i\omega \right) &=&1+\frac{\omega _{P}^{2}}{\omega \left(
\omega +\gamma \right) }  \label{epsDrude}
\end{eqnarray}
This model describes not only the plasma response of conduction electrons
with $\omega _{P}$ still interpreted as the plasma frequency but also their
relaxation, $\gamma $ being the inverse of the electronic relaxation time.

The relaxation parameter $\gamma $ is much smaller than the plasma
frequency. For Al, Au, Cu in particular, we will find in the next section
values for the ratio $\frac{\gamma }{\omega _{P}}$ of the order of $4\times
10^{-3}$. Hence relaxation plays a significant role in the modeling of the
dielectric constant only at frequencies where the latter is much larger than
unity. Stated in different words, it has to be taken into account only when
the metallic mirror behaves as a nearly perfect reflector. This suggests
that the relaxation will not have a large influence on the Casimir effect.
This qualitative argument is confirmed by the result of a numerical
integration reported on figure \ref{fig-plasma}. With a value of $\frac{%
\gamma }{\omega _{P}}$ equal to $4\times 10^{-3}$, that is of the order of
the real values obtained for Al, Au, Cu, the variation of $\eta _{F}$
remains smaller than 2\%. It thus plays a marginal role at the level of
accuracy aimed at but it is easy and safer to take it into account.

\section{Real metals}

For metals like Al, Au, Cu, the dielectric constant departs from the Drude
model when interband transitions are reached, that is when the photon energy
reaches a few $e$V. Hence, a more precise description of the dielectric
constant, taking into account the known data on optical properties of these
metals, has to be used for evaluating the force in the sub-$\mu $m range.

The dielectric response function for real frequencies may be written in
terms of real and imaginary parts $\varepsilon ^{\prime }$ and $\varepsilon
^{\prime \prime }$ obeying general causality relations 
\begin{eqnarray}
\varepsilon \left( \omega \right) &=&\varepsilon ^{\prime }\left( \omega
\right) +i\varepsilon ^{\prime \prime }\left( \omega \right)  \nonumber \\
\varepsilon ^{\prime }\left( \omega \right) -1 &=&\frac{2}{\pi }{\cal P}%
\int_{0}^{\infty }\frac{x\varepsilon ^{\prime \prime }\left( x\right) }{%
x^{2}-\omega ^{2}}{\rm d}x
\end{eqnarray}
Causality relations also allow one to obtain the dielectric constant at
imaginary frequencies $\varepsilon \left( i\omega \right) $ from the
function $\varepsilon ^{\prime \prime }\left( x\right) $ evaluated at real
frequencies $x$, that is also the oscillator strength characterizing the
material \cite{LandauECM9} 
\begin{equation}
\varepsilon \left( i\omega \right) -1=\frac{2}{\pi }\int_{0}^{\infty }\frac{%
x\varepsilon ^{\prime \prime }\left( x\right) }{x^{2}+\omega ^{2}}{\rm d}x
\label{epsIm}
\end{equation}
When discussing optical data, we will measure frequencies either in $e$V or
in rad/s, using the equivalence 1~$e$V $=1.537\times 10^{15}$~rad/s.

The values of the complex index of refraction, measured through different
optical techniques, are tabulated as a function of frequency in several
references \cite{Palik,McGrawHill,CRC98}. Optical data may vary from one
reference to another. Available data do not cover the whole frequency range
and they have to be extrapolated. These two problems may cause variations of
the results obtained for $\varepsilon \left( i\omega \right) $ and,
therefore, for the Casimir force. This is why we explain in detail how we
proceed from the input, the optical data, to the output of the process, the
reduction factors for Casimir force and energy.

\begin{figure}[tbh]
\centerline{\psfig{figure=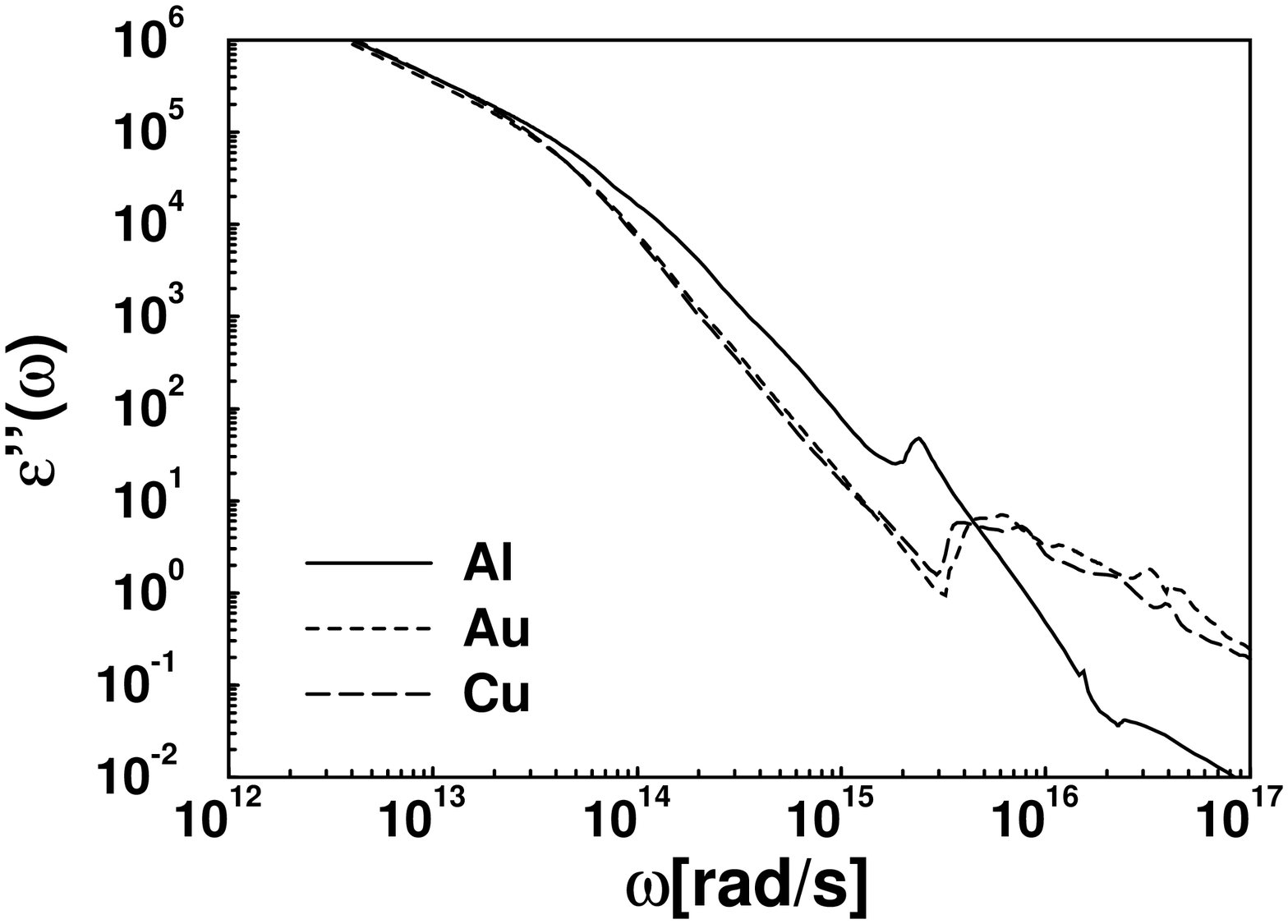,width=7cm}}
\centerline{\psfig{figure=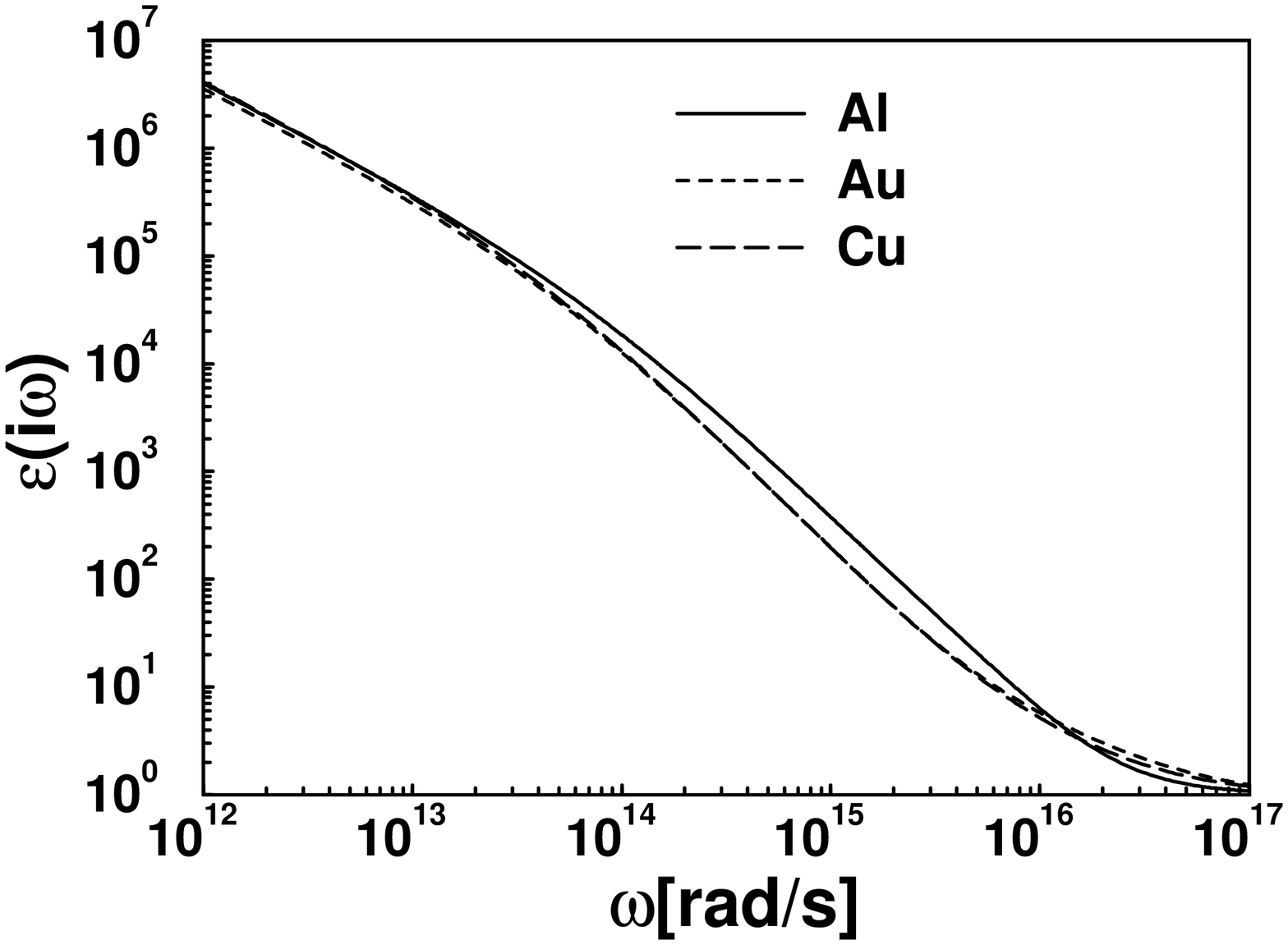,width=7cm}}
\caption{The imaginary part of the dielectric constant as a function of real
frequency (upper graph) and the dielectric constant as function of imaginary
frequency (lower graph) for Al (solid line), Au (dotted line) and Cu (dashed
line). At low frequencies the data fit a Drude model corresponding to the
contribution of conduction electrons. Peaks in the imaginary part of the 
dielectric 
function correspond to deviations from the Drude model associated with interband
transitions.}
\label{fig-eps}
\end{figure}

Figure \ref{fig-eps} shows the values for $\varepsilon ^{\prime \prime
}(\omega )$ as a function of frequency $\omega $ for the three metals Al, Au
and Cu. All data are taken from \cite{Palik,McGrawHill} with a frequency
range $0.04-1000~e$V for Al and $0.1-1000~e$V for Au and Cu. A large number
of points is available in these sources so that the interpolation between
these points does not raise any difficulty. However the data have to be
extrapolated at low frequencies to increase the domain over which the
integrations are performed. At energies around $0.1~e$V the optical
properties are quite well described by the contribution of conduction
electrons. Hence data available at these energies may be nicely fitted with
a Drude model. For Al the corresponding Drude parameters are given in \cite
{Palik} as $\omega _{P}=11.5~e$V and $\gamma =50$~m$e$V. For Au and Cu there
are not enough optical data at low frequencies to permit a determination of
the two parameters $\omega _{P}$ and $\gamma $ separately. Here we use
additional information namely the estimation of $\omega _{P}$ coming from
solid state physics \cite{AshcroftMermin,LandauECM9}. Precisely we write 
\begin{eqnarray}
\omega _{P}^{2} &=&\frac{4\pi Ne^{2}}{m^{*}}=\frac{Nq^{2}}{\varepsilon
_{0}m^{*}}  \nonumber \\
N &=&ZN_{a}
\end{eqnarray}
where $N$ is the number of conduction electrons per unit volume, that is
also the product of the number $Z$ of electrons per atom by the atomic
number density $N_{a}$, $q$ is the charge of electron and $m^{*}$ is the
effective mass of conduction electrons. This mass is different of the mass $%
m $ of free electrons. The same correction may be described as a change of
the effective number of conduction electrons per atom from $Z$ to $Z^{*}$.
We keep the former description, use $Z=1$ for Cu and Au, and choose 
for effective masses of conduction electrons the values $\frac{m^{*}}{m}%
\simeq 1$ for Au and $\frac{m^{*}}{m}\simeq 1.45$ for Cu \cite
{Schulz57,Ehrenreich62}. 
\noindent With these assumptions we obtain nearly equal values for the
plasma frequency of Au and Cu, $\omega _{P}=9.0~e$V. This corresponds to a
plasma wavelength of 136 nm for Au and Cu, to be compared with the plasma
wavelength of 107 nm for Al. Then the optical data of \cite{Palik} allow us
to deduce the relaxation parameter $\gamma $ fitting the low energy data
points with a Drude model. We obtain in this manner $\gamma =35$~m$e$V for
Au and $\gamma =30$~m$e$V for Cu. These values correspond respectively to $%
\frac{\gamma }{\omega _{P}}=3.8\times 10^{-3}$ and $\frac{\gamma }{\omega
_{P}}=3.3\times 10^{-3}$ to be compared to $\frac{\gamma }{\omega _{P}}%
=4.4\times 10^{-3}$ for Al. Note that we have given deliberately all the
numerical values in this paragraph with a limited accuracy since slightly
different values could have been obtained as well, starting from different
sources or using different criteria for choosing the values. This problem of
extrapolation of optical data at low frequencies is certainly a cause for
systematical errors in the estimation of the Casimir force.

The dielectric constant for imaginary frequencies $\varepsilon \left(
i\omega \right) $ is then obtained by numerical integration of relation (\ref
{epsIm}). Of course, the integration cannot be performed over the whole
range $[0,\infty ]$ of frequencies so that we have to give details about the
integration procedure. We are mainly interested in experimentally explored
plate separations in the range $0.1-10~\mu $m. These separations correspond
to frequencies in the range $0.1-10~e$V. We thus need reliable values for $%
\varepsilon (i\omega )$ with $\omega $ ranging from $10^{-4}$ to $10^{3}~e$%
V. To this aim we have to integrate (\ref{epsIm}) over real frequencies
covering a still broader range $10^{-6}-10^{4}~e$V. In order to test the
integration procedure we have varied the integration range by half an order
of magnitude which changed the result by less than 1 \%. The curves obtained
for the three metals are shown in the lower graph of figure \ref{fig-eps}.
In particular the curves for Au and Cu are nearly identical over the whole range 
of frequencies.

\begin{figure}[tbp]
\centerline{\psfig{figure=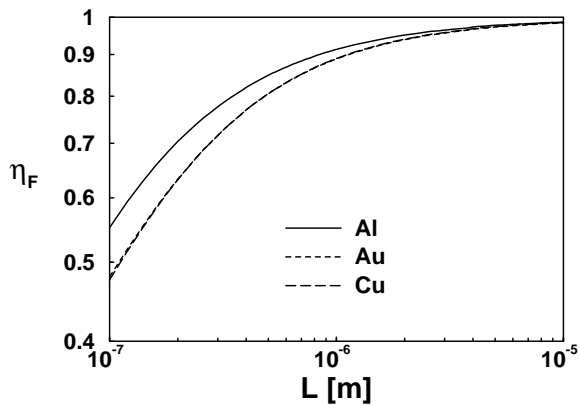,width=7cm}}
\centerline{\psfig{figure=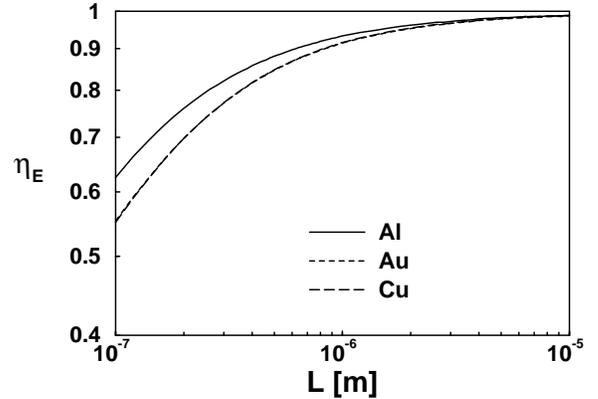,width=7cm}}
\caption{Reduction of the Casimir force (upper graph) and energy (lower
graph) between metallic reflectors with respect to perfectly reflecting
mirrors as a function of their distance $L$. The three curves correspond to
Al (solid line), Au (dotted line) and Cu (dashed line).}
\label{fig-eta}
\end{figure}

The Casimir force and energy are then calculated by numerical integration of
equations (\ref{etaF},\ref{etaE}). The integration range is chosen as $%
10^{-4}-10^{3}~e$V in order to evaluate the Casimir force for plate
separations in the range $0.1-10~\mu $m. The same test of the integration
procedure has been performed leading to an error less than 1.5\% for $\eta
_{F}$ and 2\% for $\eta _{E}$. The limit of perfect reflectors has been
reproduced with an error less than 1\%. Figure \ref{fig-eta} shows the
reduction of the Casimir force and energy between metallic mirrors with
respect to perfectly reflecting mirrors for the three metals. The force is
reduced when going from Al to Au and has nearly the same value for Au and
Cu. This directly reflects the behavior of the 
\noindent dielectric constants $%
\varepsilon (i\omega )$ which decrease along the same series on figure \ref
{fig-eps}. As already discussed in the previous section the reduction $\eta
_{E}$ is less pronounced than $\eta _{F}$.

We give in the following table a few numerical values for the reduction
factors $\eta _{F}$ and $\eta _{E}$ for the three metals at three typical
distances 
\begin{equation}
\begin{array}{cccc}
& \quad {\rm Al}\quad & \quad {\rm Au}\quad & \quad {\rm Cu}\quad \\ 
\eta _{F}\left[ 0.1~\mu {\rm m}\right] & 0.55 & 0.48 & 0.48 \\ 
\eta _{E}\left[ 0.1~\mu {\rm m}\right] & 0.63 & 0.55 & 0.55 \\ 
\eta _{F}\left[ 0.5~\mu {\rm m}\right] & 0.85 & 0.81 & 0.81 \\ 
\eta _{E}\left[ 0.5~\mu {\rm m}\right] & 0.88 & 0.85 & 0.85 \\ 
\eta _{F}\left[ 3.0~\mu {\rm m}\right] & 0.96 & 0.96 & 0.96 \\ 
\eta _{E}\left[ 3.0~\mu {\rm m}\right] & 0.97 & 0.97 & 0.97
\end{array}
\end{equation}
We remind once again that $\eta _{E}$ is the reduction factor for energy
between two plane mirrors, that is also the estimate of reduction factor for
the force in the plane-sphere geometry. From our analysis the factors $\eta
_{F}$ and $\eta _{E}$ turn out to be the same for Au and Cu. Incidentally, as 
the dielectric constants of Au and Cu are nearly the same, mirrors built with a 
layer of Au on a slab of Cu would not lead to different results.
This seems to
solve a difficulty in the analysis of experimental results of \cite
{Lamoreaux97}. The values obtained here for Al at $0.1~\mu {\rm m}$ and Cu
at $0.5~\mu {\rm m}$ correspond to those found in \cite{Lamoreaux99}. But
significant differences appear for Au at $0.5~\mu {\rm m}$ where we find
values of $\eta _{F}$ and $\eta _{E}$ exceeding by 23\% and 18\%
respectively the values given in \cite{Lamoreaux99}. Furthermore the
agreement between our result for Cu at $0.5~\mu {\rm m}$ and the one in \cite
{Lamoreaux99} appears to be an accidental crossing between two curves having
quite different behaviors as functions of distances. Since these differences
have important consequences for the comparison of experimental results with
theory, we discuss them in detail in Appendix \ref{app-Copper}. 

\section{Conclusion}

The Casimir force has now been experimentally explored at distances in the 
sub-$\mu $m range and the reduction of the force due to
finite conductivity of metals has been observed. For an accurate comparison of 
the experimental
results with theory, it is necessary to dispose of precise
theoretical expectations.

In this paper we have presented a detailed analysis of the influence of the
imperfect reflection on the Casimir force between two plane metallic plates.
In particular, we have given a precise evaluation of the reduction factor
for metals used in recent experiments, that is Al, Au and Cu. This factor
becomes significant at distances smaller than $1~\mu $m and it reaches
values of about 50\% for the smallest explored distances.

The reduction factor $\eta _{E}$ calculated in the present paper for the
energy between plane plates can be used to infer the reduction factor for
the force in the plane-sphere geometry if the proximity theorem is trusted.
However the accuracy of this theorem is not known. Other corrections have
also to be taken into account. Thermal corrections are significant at
distances larger than a few $\mu $m but have not been seen in the experiment
where these distances were explored \cite{Lamoreaux97}. The roughness
corrections are also expected to play an important role \cite
{Bezerra97,Roy99PRL}.

In these conditions it is premature to claim that a good agreement has been
reached between experiments and theory. It is worth developing new
experiments using either the same techniques or different ones \cite
{Onofrio95,Grado99}. More work is also needed on the theoretical side, in
particular for obtaining more reliable estimations of the effect of geometry
and roughness on the Casimir force. Such efforts are certainly worthwhile
not only because of the interest of reaching conclusions on the Casimir
force but also for making it possible to control its effect when studying
small short range forces \cite{Fischbach,Carugno97,Long99,Bordag99}.

\noindent {\bf Acknowledgements} We wish to thank Ephraim Fischbach,
Marc-Thierry Jaekel, David Koltick, Vladimir Mostepanenko and Roberto Onofrio 
for stimulating discussions.

\appendix

\section{The plasma corrections}

\label{app-Plasma}In this appendix we compare the reduction factor $\eta
_{F} $ evaluated for the Casimir force from expressions (\ref{etaF},\ref
{rThick}) using the plasma model (\ref{epsPlasma}) with different
approximants which have been used to discuss recent experimental results
(see for example \cite{Lamoreaux99,Klimchitskaya99,Bezerra97}).

\begin{figure}[tbp]
\centerline{\psfig{figure=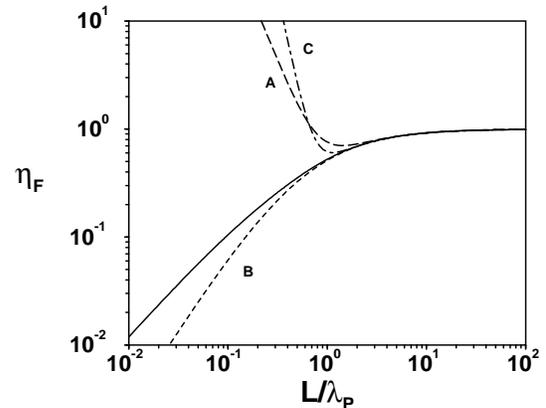,width=7cm}}
\caption{Reduction factor $\eta _{F}$ for the Casimir force as a function of
the ratio $\frac{L}{\lambda _{P}}$ when the finite conductivity is described
by a plasma model (solid line). This curve is compared to different
approximants (dashed lines) used in the literature and described in the
text. These approximants are reasonable at large distances but depart from
the exact result when $\frac{L}{\lambda _{P}}$ approaches unity.}
\label{fig-approx}
\end{figure}

The exact result is the solid line of figure \ref{fig-plasma} reproduced as
the solid line on figure \ref{fig-approx}. Its behavior at long distances $%
\left( L\gg \lambda _{P}\right) $ corresponds to the known development \cite
{Mostepanenko85} 
\begin{equation}
\eta _{A}=1-\frac{8}{3\pi }\frac{\lambda _{P}}{L}+\frac{6}{\pi ^{2}}\left( 
\frac{\lambda _{P}}{L}\right) ^{2}
\end{equation}
drawn as long dashes A on figure \ref{fig-approx}. Another interpolation
formula may be deduced from this behavior as \cite{Mostepanenko85} 
\begin{equation}
\eta _{B}=\left( 1+\frac{11}{6\pi }\frac{\lambda _{P}}{L}\right) ^{-\frac{16%
}{11}}
\end{equation}
This formula, drawn as short dashes B on figure \ref{fig-approx}, presents
the advantage of being positive at all distances and also being a monotonic
function of distance, two important features of the exact result. However it
fails to reproduce the asymptotic variation of $\eta _{F}$ at small
distances (compare with (\ref{etashortF})). Another approximant, obtained by
developing $\eta _{B}$ at the fourth order in $\frac{\lambda _{P}}{L}$, has
sometimes been used \cite{Klimchitskaya99} 
\begin{eqnarray}
\eta _{C}&=& \eta_A -\frac{38}{3\pi ^{3}}\left( \frac{\lambda _{P}}{L}%
\right) ^{3} + \frac{931}{9\pi ^{4}}\left( \frac{\lambda _{P}}{L}\right) ^{4}
\end{eqnarray}
It is drawn as the dotted-dashed line C on figure \ref{fig-approx}. On the
whole, figure \ref{fig-approx} clearly shows that all these approximants
fail to reproduce the correct behavior at distances smaller than the plasma
wavelength $\lambda _{P}$.

Incidentally an interesting approximant may be defined by the following
formula 
\begin{equation}
\eta _{U}=\frac{1}{1+\frac{8}{3\pi }\frac{\lambda _{P}}{L}}
\end{equation}
It fits the known behavior of the exact result at the first-order, but not
the second-order one, in $\frac{\lambda _{P}}{L}$. It is proportional to $%
\frac{L}{\lambda _{P}}$ at small distances as the correct result (\ref
{etashortF}). Moreover the coefficient $\frac{3\pi }{8}$ has a value $1.178$
very close to the proportionality coefficient $\alpha \simeq 1.193$
appearing in (\ref{etashortF}), the relative difference being of the order
of 1\%. Hence, $\eta _{U}$ may be considered as a uniform approximant
reproducing the variation of $\eta _{F}$ over the whole range of distances. It
reproduces the exact result everywhere with an error of at most 5\%. This
precision is however not sufficient for it to be used in the place of the
correct result.

\section{The case of Copper}

\label{app-Copper}In this appendix, we compare the reduction factors $\eta
_{F}$ and $\eta _{E}$ obtained for Copper from the computations of the
present paper and those of \cite{Lamoreaux99}. Both derivations are based on
the same procedure which we have already described in the text. Optical data
taken from different references agree reasonably well between each other. We
however point out that quite different techniques are used here and in \cite
{Lamoreaux99} for interpolating between available data and extrapolating at
low frequencies and that these differences are responsible for significant
deviations in the behaviors of $\eta _{F}$ and $\eta _{E}$ as functions of
the plate separation.

\begin{figure}[tbh]
\centerline{\psfig{figure=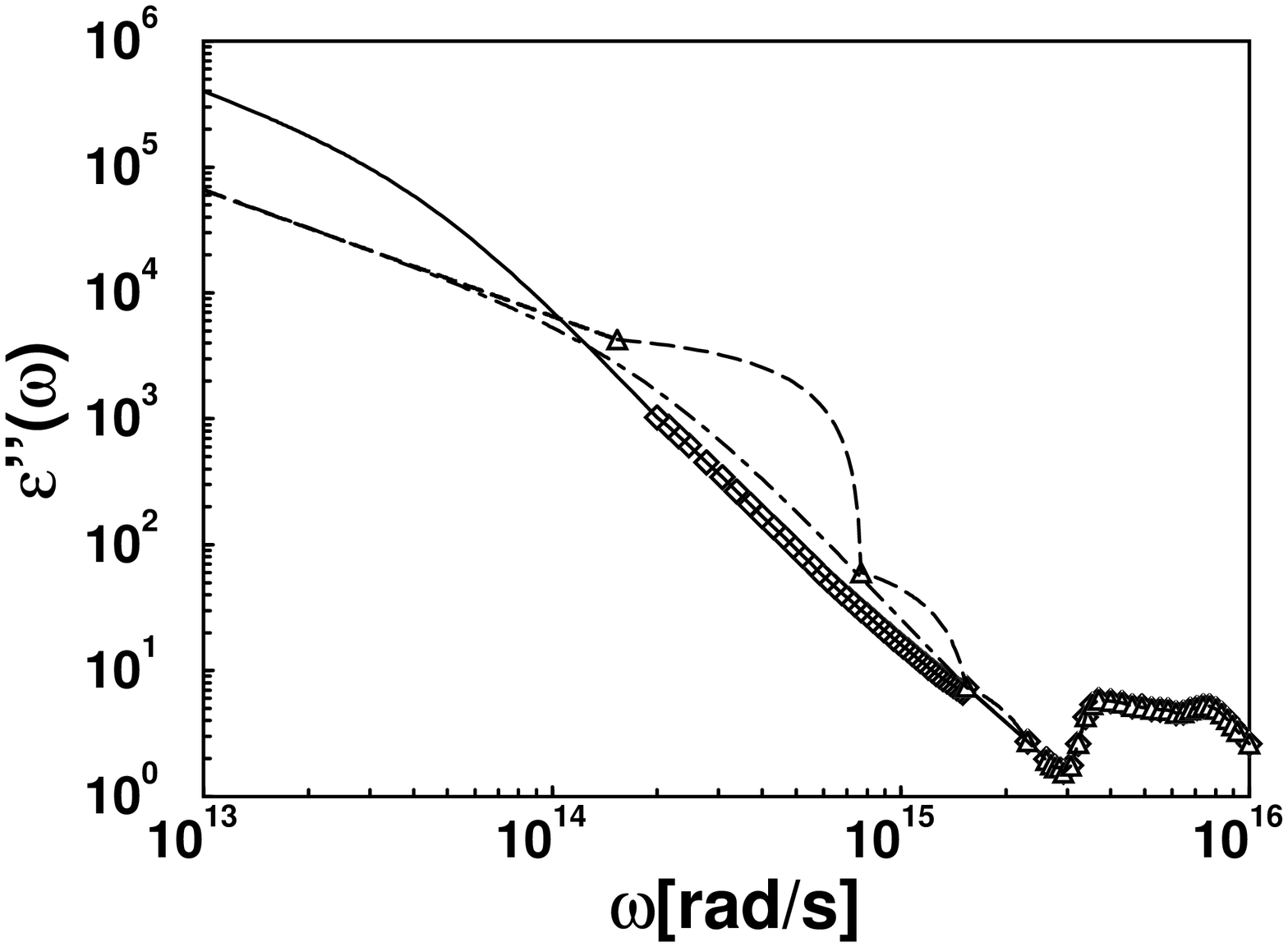,width=7cm}}
\centerline{\psfig{figure=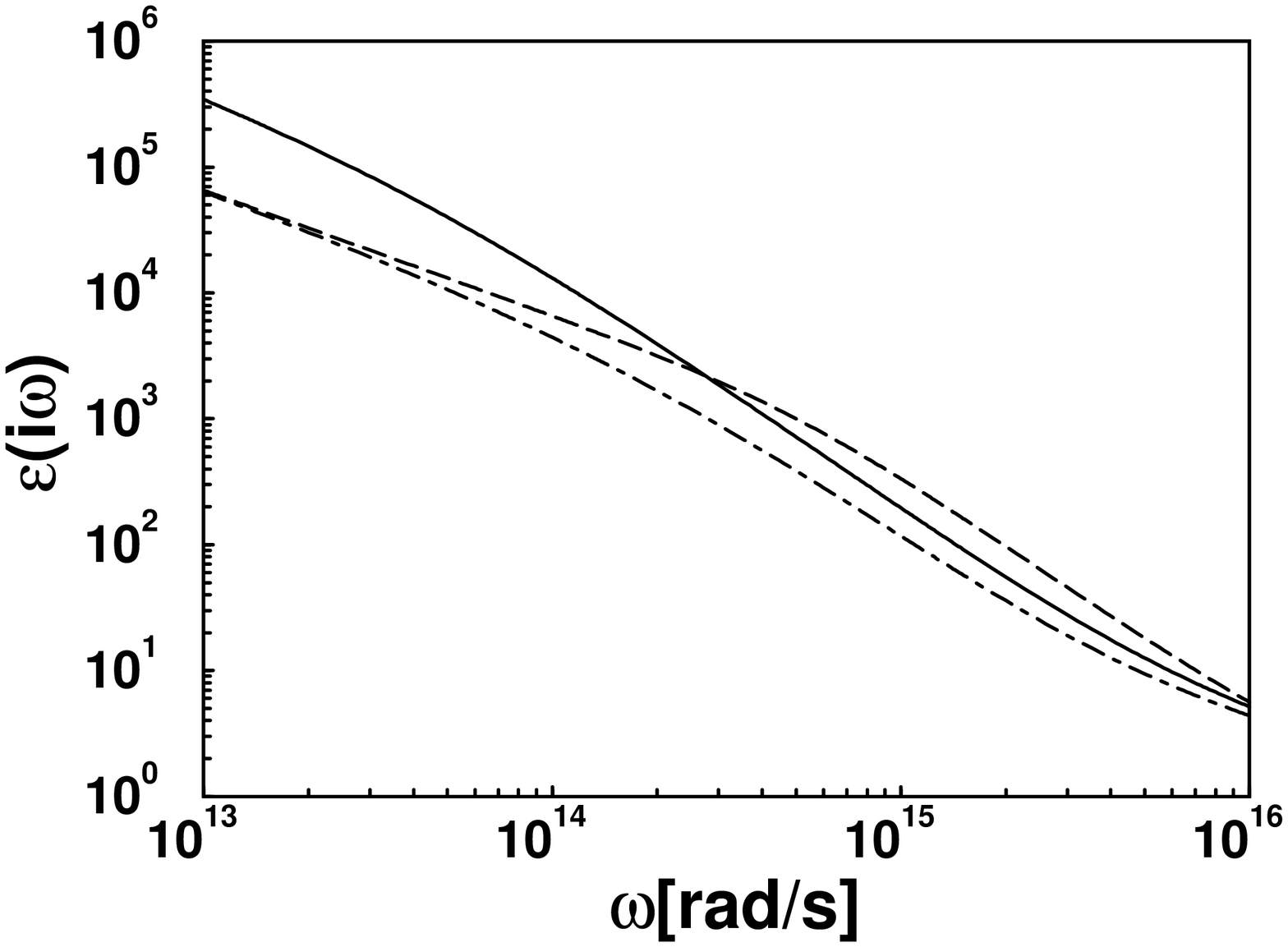,width=7cm}}
\caption{Imaginary part of the dielectric constant as a function of
frequency (upper graph) and dielectric constant as function of imaginary
frequency (lower graph) for Copper. The diamonds are data points from reference 
\protect\cite{Palik,McGrawHill} and the solid line is our extrapolation at
low frequencies of these data. The triangles correspond to data points from 
\protect\cite{CRC98}. The dashed line represents the extrapolation at low
frequencies and the interpolation between the last three data points as in 
\protect\cite{Lamoreaux99}. The dotted-dashed line corresponds to another
extrapolation of the same data. More detailed explanations are given in the
text.}
\label{fig-epsCu}
\end{figure}

The upper graph of figure \ref{fig-epsCu} shows three different plots of the
imaginary part of the dielectric constant. The first one is the one
explained in the present paper with data points taken from \cite
{Palik,McGrawHill} (diamonds) and extrapolation at low frequency with a
Drude model (solid line on figure \ref{fig-epsCu}). Our procedure is
explained in more detail in the main part of this paper. The corresponding
Drude parameters, a plasma frequency $\omega _{P}=9.0~e$V and a relaxation
parameter $\gamma =30$~m$e$V are in reasonable agreement with existing
knowledge from solid state physics.

The second plot has been designed by ourselves as an attempt to reproduce
the computations of \cite{Lamoreaux99}. The triangles are optical data taken
from \cite{CRC98}. These data are not exactly identical but they are in
reasonable agreement with those taken from \cite{Palik,McGrawHill}. However
only three data points are given in \cite{CRC98} for the frequency range $%
10^{14}$-$2\times 10^{15}$ rad/s whereas a much larger number of data points
may be found in \cite{Palik,McGrawHill}. In contrast to our treatment, a
specific interpolation procedure had therefore to be used in \cite
{Lamoreaux99} to fill the gaps between the data points. Although this
procedure is not described explicitly in \cite{Lamoreaux99} we have been
able to reproduce a curve having the same appearance (compare with figure 1a
in \cite{Lamoreaux99}). This curve results from a linear interpolation
between the data points on a lin-lin scale. It appears clearly on figure \ref
{fig-epsCu} that this interpolation procedure produces bumps on the
dielectric response functions (dashed line) which are largely outside the
data known from \cite{Palik,McGrawHill}. Optical data are in fact consistent
with a linear interpolation on a log-log scale rather than on a lin-lin
scale. This is the first important difference between the two treatments.
The second important difference is associated with the extrapolation of data
at low frequencies. In \cite{Lamoreaux99} the data points were extrapolated
by a power law proportional to $1/\omega $ starting from the lowest
frequency data available in \cite{CRC98}. The whole curve of \cite
{Lamoreaux99} is not at all consistent with a Drude model at frequencies
below $10^{15}$ rad/s. To summarize this presentation of the dielectric
function used in \cite{Lamoreaux99} we may say that it corresponds to values
too large in the range $10^{14}$-$2\times 10^{15}$ rad/s by a factor which
can be more than $10$ and too small below $10^{14}$ rad/s by a factor up to $6$.

The case of low frequency extrapolation requires more cautious discussions.
As explained in the main part of this paper, the optical data available for
Cu do not permit an unambiguous estimation of the two parameters $\omega
_{P} $ and $\gamma $ separately. Other couples of value can be chosen which
would also be consistent with optical data. To make this point explicit, we
have drawn a third plot on figure \ref{fig-epsCu} (dashed-dotted line) which
corresponds to a Drude model fitting the optical data of \cite{CRC98} and
the low frequency behavior of \cite{Lamoreaux99}. Obviously it does not
reproduce the extra bumps of \cite{Lamoreaux99}. The associated Drude
paramaters $\omega _{P}=7.5~e$V and $\gamma =130$~m$e$V correspond to an
effective mass $\frac{m^{*}}{m}\simeq 2.1$ and to a ratio $\frac{\gamma }{%
\omega _{P}}\simeq 1.7\times 10^{-2}$ which are quite different from those
used in our treatment. In order to have an indication of the effect of the
uncertainties associated with optical data, we will however proceed to the
computations with this curve, too.

\begin{figure}[tbh]
\centerline{\psfig{figure=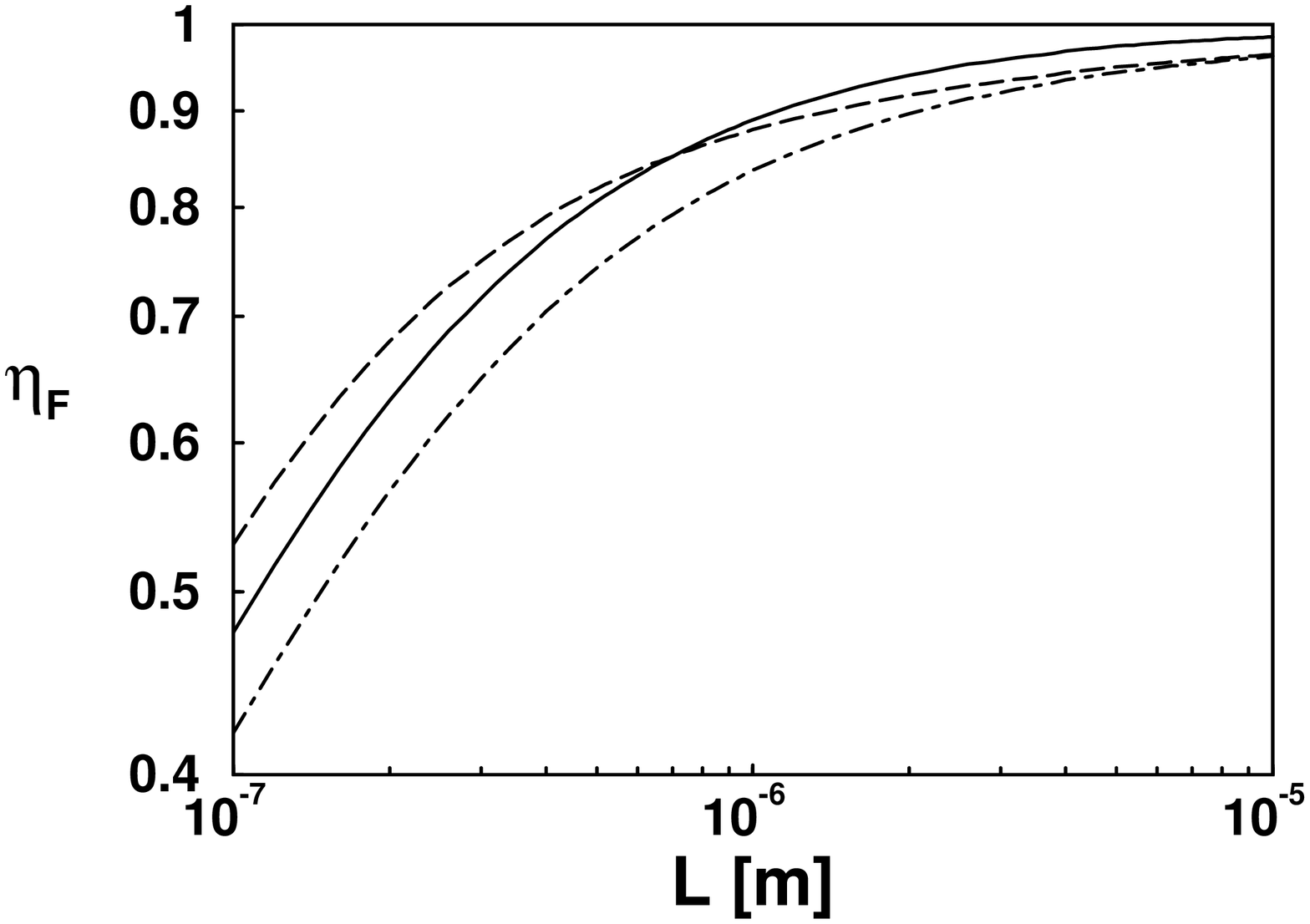,width=7cm}}
\centerline{\psfig{figure=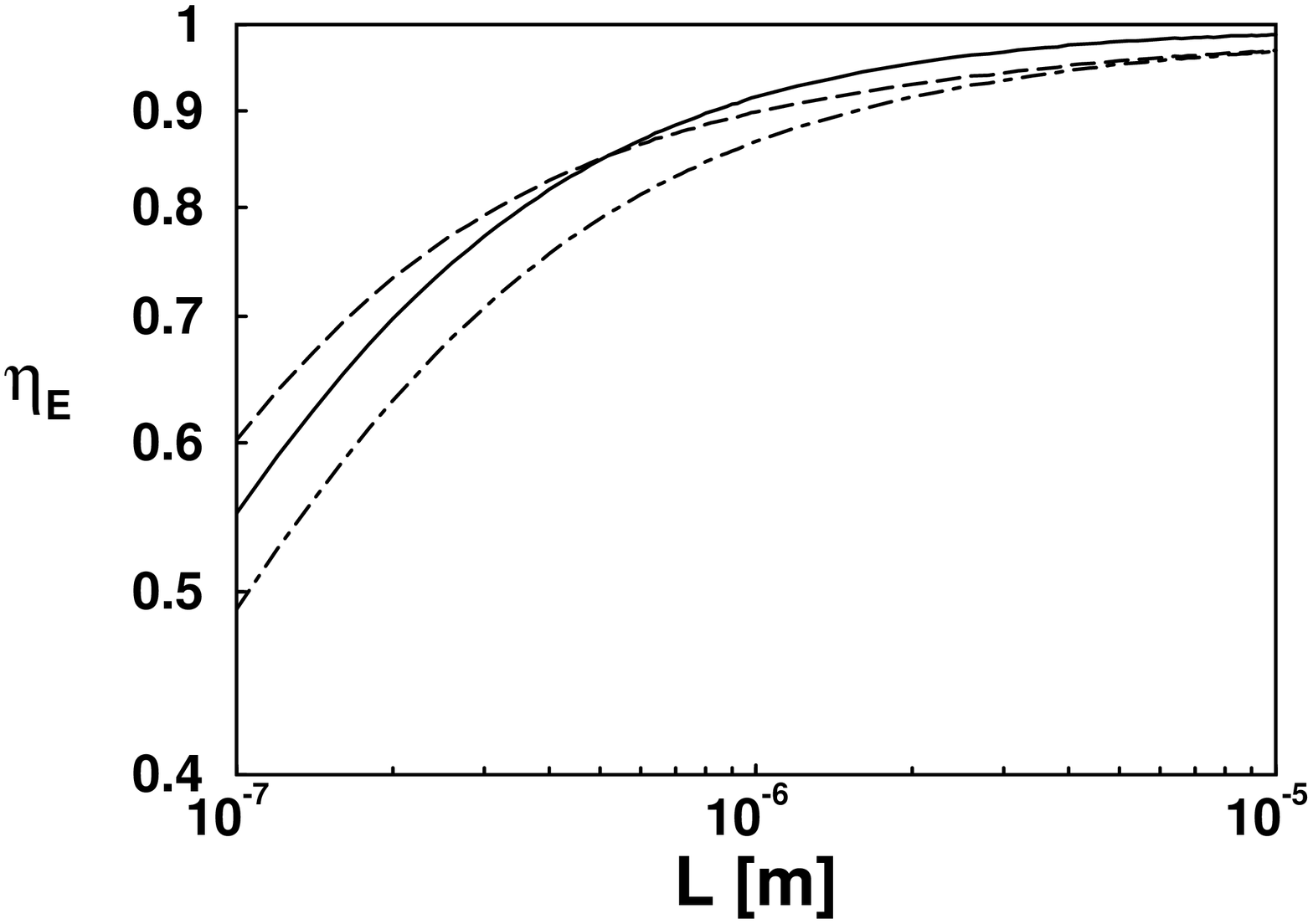,width=7cm}}
\caption{Reduction factors $\eta_F$ for the Casimir force (upper graph) and $%
\eta_E$ for the Casimir energy (lower graph) as a function of the plate
separation $L$. The solid line corresponds to the calculation of the present 
paper, 
the dashed line to the calculation in \protect\cite{Lamoreaux99}. The 
dotted-dashed
line corresponds to a calculation with a Drude model fitting the optical data of 
\protect\cite{CRC98} and the low frequency behavior of 
\protect\cite{Lamoreaux99}.}
\label{fig-etaCu}
\end{figure}

We now perform the calculations as explained in the main part of the text.
The lower graph on figure \ref{fig-epsCu} shows the different results for
the dielectric constant $\varepsilon (i\omega )$ as a function of imaginary
frequency for the three different dielectric functions. As expected from the
previous discussion, the values $\varepsilon (i\omega )$ found in \cite
{Lamoreaux99} are too small at frequencies lower than $10^{14}$ rad/s but
too large around $10^{15}$ rad/s when compared to those deduced from our
calculation. This has a significant consequence for the evaluation of the
reduction factors $\eta _{F}$ and $\eta _{E}$ drawn on figure \ref{fig-etaCu}
for a plate separation ranging from $0.1~\mu $m to $10~\mu $m.

Our reconstruction of the computations of \cite{Lamoreaux99} reproduce
pretty well the published results at the distance of $0.5~\mu {\rm m}$ for
which numerical values are given. However, the general behaviors of the
curves are quite different. The Casimir force and Casimir energy obtained
from the optical data used in \cite{Lamoreaux99} are too large at small
distances and too small at large distances. These features, which are made
explicit with values of $\eta _{E}$ given in the following table, are
consistent with the discussions of the preceding paragraphs. The three
columns 1, 2, 3 correspond respectively to the solid lines, dashed lines and
dotted-dashed lines of figures \ref{fig-epsCu}-\ref{fig-etaCu} 
\begin{equation}
\begin{array}{cccc}
& \quad {\rm 1}\quad & \quad {\rm 2}\quad & \quad {\rm 3}\quad \\ 
\eta _{E}\left[ 0.1~\mu {\rm m}\right] & 0.55 & 0.60 & 0.49 \\ 
\eta _{E}\left[ 0.5~\mu {\rm m}\right] & 0.85 & 0.85 & 0.79 \\ 
\eta _{E}\left[ 3.0~\mu {\rm m}\right] & 0.97 & 0.94 & 0.93
\end{array}
\end{equation}
The crossing of the results in columns 1 and 2 at the distance of $0.5~\mu $%
m appears as an accidental compensation of these two flaws. The relative
difference between the two results may be as large as 10\%.

A claim of agreement between experiment and theory could be based on a
comparison of values of $\eta _{E}$ obtained at different distances with
values in the column 1 or, perhaps, in the column 3. As explained above,
both columns correspond to reasonable extrapolations of the optical data.
The advantage of column 1 over column 3 lies in values of the Drude parameters
in better accordance with the knowledge in solid state physics. The
difference between columns 1 and 3 may be considered as giving an idea of
the uncertainties associated with the incompleteness of optical data. In any
case the two corresponding curves, drawn as solid and dotted-dashed lines on
figure \ref{fig-etaCu}, have similar dependences on the plate separation
although the absolute values are shifted from one curve to the other. In
contrast the dashed curve on figure \ref{fig-etaCu} which corresponds to the
calculations of \cite{Lamoreaux99} and crosses the two former curves cannot
be considered as consistent with the known optical data.

\end{document}